\documentclass{ws-ijmpc}

\usepackage{epsfig}
\usepackage{latexsym,amsmath}

\newcommand{\cats}[1]{\ensuremath{\big|#1\big>}}
\newcommand{\bras}[1]{\ensuremath{\big<#1\big|}}
\newcommand{\basis}[1]{\ensuremath{\big\{\cats{#1}\big\}}}

\newcommand{\tm}[4]{\ensuremath{\bras{{#1}{#2}}\tau\cats{{#3}{#4}}}}

\newcommand{\Op}{\ensuremath{\mathcal{O}}}

\begin{document}

\markboth{Erik Bartel, Andreas Schadschneider}
{Quantum Corner-Transfer Matrix DMRG}

\catchline{}{}{}{}{}

\title{QUANTUM CORNER-TRANSFER MATRIX DMRG}

\author{ERIK BARTEL \and ANDREAS SCHADSCHNEIDER}

\address{Institut f\"ur Theoretische Physik, Universit\"at zu K\"oln\\
50937 K\"oln, Germany\\
as@thp.uni-koeln.de}

\maketitle

\begin{history}
\received{\today}
\revised{Day Month Year}
\end{history}

\begin{abstract}
We propose a new method for the
  calculation of thermodynamic properties of one-dimensional quantum
  systems by combining the TMRG approach with the corner
  transfer-matrix method.  The corner transfer-matrix DMRG method
  brings reasonable advantage over TMRG for classical systems.  We
  have modified the concept for the calculation of thermal properties
  of one-dimensional quantum systems.  The novel QCTMRG algorithm is
  implemented and used to study two simple test cases, the classical
  Ising chain and the isotropic Heisenberg model.  In a discussion,
  the advantages and challenges are illuminated.
\end{abstract}

\ccode{PACS Nos.: {02.70.-c}, {64.60.De}, {05.70.-a}, {05.10.Cc}}

\section{Introduction}\label{sec:Intro}

The Density Matrix Renormalization Group (DMRG) \cite{W92,W93} as a
major technique for one-dimensional quantum systems provides a
variational method in the space of matrix-product states
\cite{OR95,VGCi04}.  The key idea of DMRG comprises the repeated basis
truncation by density matrix projection in an iteratively enlarged
system.  This idea turned out to be extraordinary successful beyond
the original purpose of computing the low-energy
spectrum of quantum chains with short-range interactions
\cite{DMRG,Scho05}.

Finite-temperature properties of quantum chains can
be obtained by mapping the one-dimensional quantum system to a
two-dimensional classical lattice via Trotter-Suzuki decomposition
\cite{T59,Suzu76,S85.1,S85.2,Suz86}.  The Transfer Matrix DMRG method
(TMRG) \cite{BXG96,WX97,Shib97} yields the finite temperature
properties of infinite-sized quantum chains by applying a DMRG
algorithm to the transfer matrix of such a corresponding
two-dimensional lattice in order to find the highest contributing
eigenvalues.

In this paper, we propose a different method: The {\em Quantum Corner
Transfer Matrix DMRG (QCTMRG)} for the calculation of thermodynamic
properties of finite one-dimensional quantum systems is a variant of
Nishino's Corner Transfer Matrix DMRG (CTMRG) for classical systems
\cite{NO96}.  Here, both calculating the partition sum and obtaining
the reduced density matrix is just one step and does not involve
finding eigenvectors of large transfer matrices.  Thus, CTMRG performs
drastically better than the TMRG method for classical systems.  Can we
benefit from the ideas of the CTMRG in the quantum case?

In contrast to the classical models treated with CTMRG so
far, two substantial differences arise in dealing with a Trotter
decomposition:
\begin{itemize}
\item A significant {\em anisotropy} due to the existence of a
  well-distinguished real space and a Trotter direction.
\item Calculation of the trace in order to obtain the partition
  function demands {\em periodic boundary conditions} of the
  two-dimensional plane.
\end{itemize}
Both aspects will be considered our QCTMRG approach.  In this paper,
we present the QCTMRG algorithm and its essential features and close
with first results and a discussion of the method's performance.

\section{CTMRG for quantum systems}\label{sec:QCTMRG_method}

\subsection{Trotter decomposition}\label{subsec:Trotter}

We consider one-dimensional quantum chains of length $L$ where the
Hamiltonian $H = \sum_{j=1}^L h_{j,j+1}$ only includes nearest-neighbor
interactions.  Generically, the partition function of
such a quantum chain does not factorize because the neighboring
interaction terms do not commute.  This difficulty has been overcome
by the Trotter-Suzuki decomposition \cite{Suzu76} which maps the
partition function of the quantum chain onto a partition function of a
two-dimensional classical chequerboard model.  A variant was
introduced by Sirker and Kl{\"u}mper \cite{SiKl02a} where the
partition function
\begin{eqnarray}
  \label{e:Sirker}
  Z = \operatorname{tr} e^{-\beta H}
   = \lim_{M \to \infty}
   \operatorname{tr}  \left[ T_{\text{R}} e^{-\epsilon H}
       T_{\text{L}} e^{-\epsilon H} \right]^{M/2}
\end{eqnarray}
where $\epsilon=\beta/M$, is decomposed into a trace over a product of
imaginary-time propagators.  Here, $T_{\text{R}}$ and $T_{\text{L}}$
are left and right shift operators.  Note that because of
translational invariance, $[T_{\text{R},\text{L}},H]=0$.  The
decomposition with shift operators overcomes disadvantages of the
common chequerboard decomposition because the spatial periodicity of
the lattice is one site \cite{SiKl02a}.  Yet, the QCTMRG algorithm as
described in the subsequent section can be adapted to the chequerboard
decomposition with only minor changes.

As a benefit from the factorization,
we are able to insert identities of the form
\begin{equation}
  \sum \cats{s_1 \dots s_L}\bras{s_1 \dots s_L}=1
\end{equation}
pictured as slices of discrete temperature or imaginary time, where the chain
state $\cats{s_1 \dots s_L}$ is the tensor product of local quantum states
$\cats{s_j}$ at site $j$. Thus, we achieve a classical two dimensional model,
spanned by the real space or chain direction and the auxiliary introduced
imaginary time or so called Trotter direction.
\begin{figure}[htbp]
  \begin{center}
      \includegraphics[width=0.2\columnwidth]{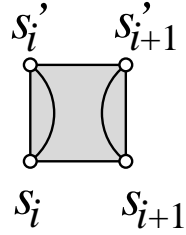}
  \end{center}
\caption{\label{f:tau}%
  Graphical representation of 4-spin transfer-matrix $\tau$.  The
  bended curves illustrate the symmetries of the object.  }
\end{figure}
The local transfer matrix (see figure \ref{f:tau}) 
\begin{equation}\label{eq:local_tau}
\bras{{s'\!\!}_{i} {s'\!\!}_{i+1}} \tau \cats{s_{i} s_{i+1}} := 
\bras{{s'\!\!}_{i} {s'\!\!}_{i+1}}
e^{-\epsilon h_{i,i+1}} \cats{s_{i} s_{i+1}}
\end{equation}
is a fourth order tensor of dimension $n \times n \times n \times n$
where $n$ is the number of states per site.
With this, we find
\begin{eqnarray}
  \bras{{s'\!\!}_1 \dots {s'\!\!}_L} e^{-\epsilon H}
  \cats{s_1 \dots s_L}
=  \sum_{\tiny
      \begin{array}{c}
       \scriptscriptstyle \{\mu_j\}\\
       \scriptscriptstyle j=1\dots L
      \end{array}} \prod_i^L
  \tm{\mu_i}{{s'\!\!}_{i+1}}{s_i}{\mu_{i+1}}
=  \sum_{\tiny
      \begin{array}{c}
        \scriptscriptstyle\{\mu_j\}\\
       \scriptscriptstyle j=1\dots L
      \end{array}} \prod_i^L
  \tm{{s'\!\!}_{i-1}}{\mu_{i}}{\mu_{i-1}}{s_{i}}.\qquad
\end{eqnarray}
With the action of the right and left shift operators
\begin{eqnarray}
  \bras{{s'\!\!}_1 \dots {s'\!\!}_L} T_{\text{R,L}} 
  \cats{s_1 \dots s_L}
  = \prod_i^L \delta_{{s'\!\!}_{i\pm1},s_{i}},
\end{eqnarray}
we get the partition function
\begin{eqnarray}%
  \label{e:Sirker2}
  Z &=&\sum_{\tiny
      \begin{array}{c}
        \{\mu^m_l\},\{s^m_l\}\\
        l=1\dots L\\
        m=1\dots M
      \end{array}}
       \prod_i^L \prod_k^{M/2}
       \tm{\mu^{2k+1}_i}{s^{2k+2}_{i}}{s^{2k+1}_i}{\mu^{2k+1}_{i+1}}
       \tm{s^{2k+1}_{i}}{\mu^{2k}_{i}}{\mu^{2k}_{i-1}}{s^{2k}_i}
\end{eqnarray}
of a two-dimensional classical lattice.  For finite $M$ the
corrections to the approximated partitions functions and free energies
have been shown to be no larger than $\mathcal{O}(\epsilon^2)$
\cite{Sir02}.
\begin{figure}[htbp]
  \begin{center}
    \includegraphics[width=0.6\columnwidth]{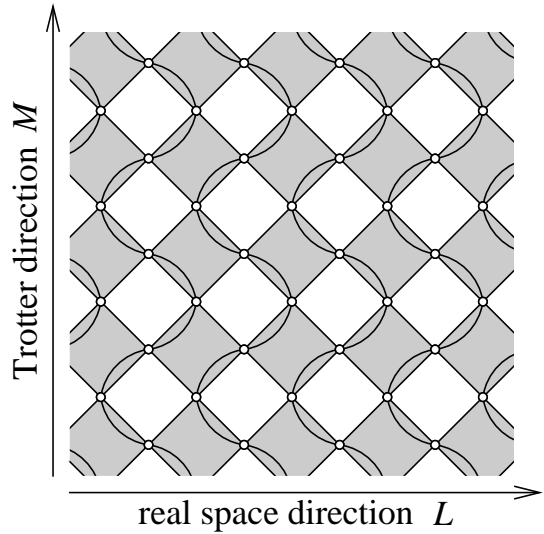}
  \end{center}
\caption{Graphical representation of the Trotter decomposition 
(\ref{e:Sirker2}).}
\label{f:Trotter-Sirker}
\end{figure}
Graphically (see figure \ref{f:Trotter-Sirker}), the decomposition is
represented as rows of alternately rotated local transfer matrices
$\tau$.

Thermal expectation values of local operators can be calculated by a
modified partition function.  We recall the statistical definition of
the ther\-mo\-dy\-na\-mi\-cal ex\-pec\-ta\-tion value
\begin{equation}\label{e:ZO}
  \langle\Op_i\rangle = \frac{Z(\Op_i)}{Z}
  \qquad\text{ with }\quad Z(\Op_i) = \operatorname{tr} \Op_i e^{- \beta H}
\end{equation}
of an operator $\Op$ at a site $i$.  In terms of a two-dimensional
classical model, the non-normalized expectation value $Z(\hat{O_i})$
is formed by replacing one standard transfer-matrix within the product
(\ref{e:Sirker2}) by the modified local transfer-matrix $\tilde{\tau}
:= \Op_i e^{-\epsilon h}$.

\subsection{Constituting tiles}\label{subsec:Tiles}

We consider the partition function of a Trotter decomposition
(\ref{e:Sirker2}).  Starting with an initial lattice small enough to
be treated analytically, our aim is to iteratively expand this system
in both directions to large enough sizes.  Periodic boundary
conditions correspond to the trace in the calculation of the partition
function.  They are physically essential in Trotter direction.  We
are, however, free to choose open boundary conditions in real space
direction.  This choice significantly reduces computational effort
because the tensor-dimensionality of the corner tiles is now three
rather than four in the case of real-space periodic boundary
conditions.  Thus, free sites at both real space edges will be
integrated out, whereas the edge states regarding Trotter direction
are left as degrees of freedom in order to permit periodical closing.

\begin{figure}[htbp]
  \begin{center}
    \includegraphics[width=0.95\columnwidth]{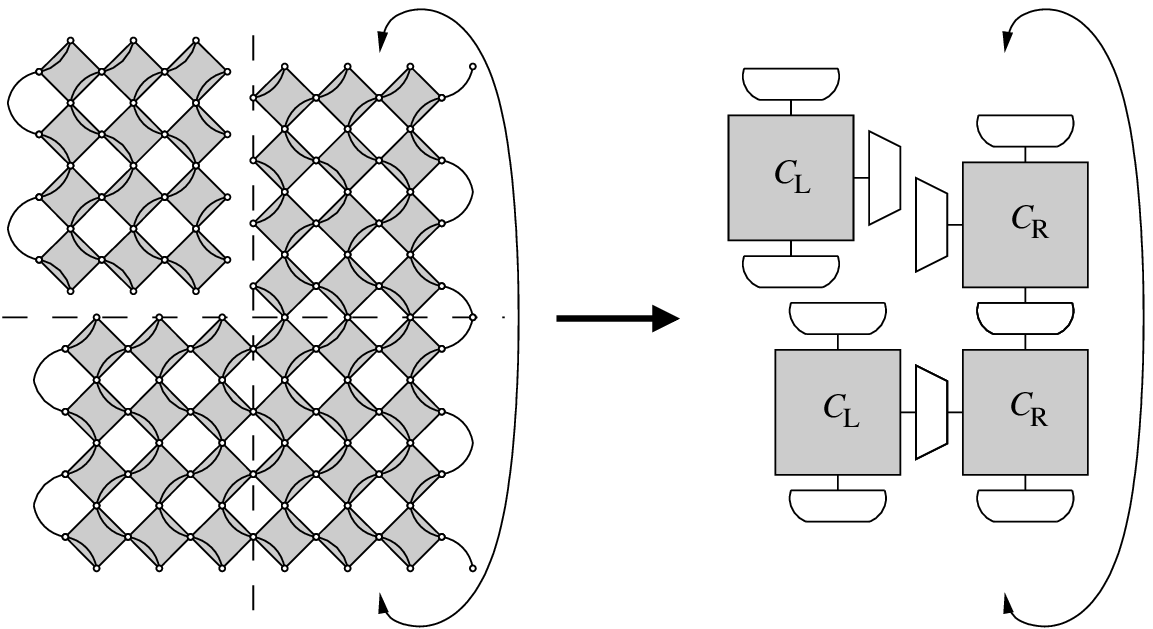}
  \end{center}
\caption{Trotter decomposition segmented into four corner tiles.
  The periodical boundary conditions are illustrated by the
  double-headed arrow.  The corner tiles $C_L,C_R$ have been depicted
  by special symbols reflecting the underlying symmetry.}
\label{f:c_tiles}
\end{figure}

At an arbitrary renormalization step, we consider a lattice of fixed
size in both directions divided into four parts (see
figure~\ref{f:c_tiles}).  Assuming free edges in Trotter direction, we
no longer deal with two-dimensional matrices as in conventional CTMRG,
but with three-dimensional tensors.  So the pieces of the system are
third and fourth order tensors and, thus, shall rather be called tiles
than--somewhat misleadingly--matrices in this context.  Here, a
composition of left and right corner tiles $C_L, C_R$ makes up the
whole system.
\begin{figure}[htbp]
  \begin{center}
    \includegraphics[width=0.95\columnwidth]{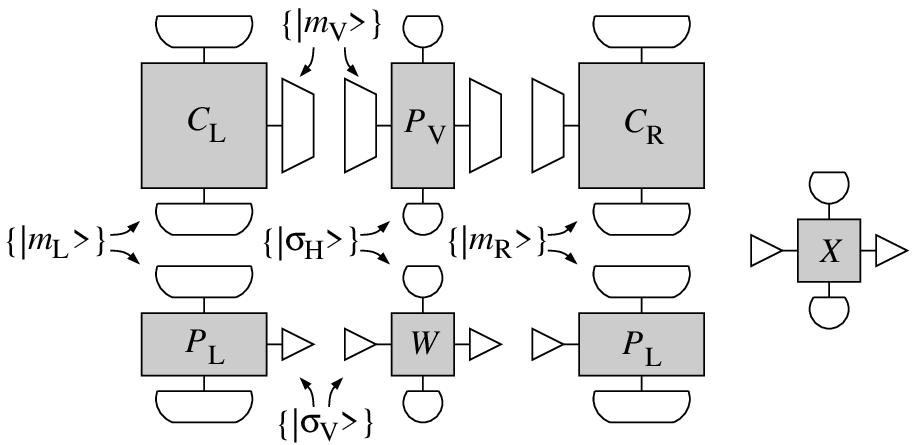}
  \end{center}
\caption{Graphical representation of left and right corner tiles $C_L,C_R$, 
  row-to-row tiles $P_L,P_R$, column-to-column tile $P_V$ and basic
  tile $W$. An additional tile $X$ is used for the measurement of
  expectation values.}
\label{f:all_tiles}
\end{figure}

Now we introduce some additional tiles which are required for
enlarging the system in the renormalization step
(figure~\ref{f:all_tiles}).  The unit cell of the two-dimensional
Trotter decomposition forms the smallest basic tile $W$ with four free
edges.  Arising from the spatial symmetries, three tiles play the role
of the row-to-row transfer matrix of CTMRG.  Here, we have for both
left and right side the row-to-row tiles $P_L$ and $P_R$ and a single
column-to-column tile $P_V$, which are tensors of third resp.\ fourth
order\footnote{Depending of the symmetries of the spin chain and the
  Trotter decomposition, $C_L$ and $C_R$ as well as $P_L$ and $P_R$
  can be represented by a single tile in special cases. Here, we
  assume the general case with no vertical reflection symmetry.}.

Together with these tiles, we have to keep account of the different
bases associated with their edges.  The basic tile $W$ demands a
vertical basis $\{\cats{\sigma_V}\}$ at the left and right edge and a
horizontal basis $\{\cats{\sigma_H}\}$ at the upper and lower edge,
both consisting of one- or two-site states depending on the size of
the unit cell of the Trotter decomposition.  These bases will be
unaffected by changes within the renormalization procedure.  The edge
bases of the corner tiles, however, are iteratively renormalized
during the CTMRG-iterations
to keep a fixed maximum size.  The maximum size of the vertical-edge
basis $\{\cats{m_V}\}$ of the corner tiles is $m_V$.  The horizontal
edges of the corner tiles are different for the left/right corner
tile.  Their bases $\{\cats{m_{L}}\},\{\cats{m_{R}}\}$ have a maximum
size of $m_L$ resp.\ $m_R$.  Tiles $P_L, P_R$ and $P_V$ include bases
which stay unaltered as well as renormalized bases corresponding to
their different edges.

\subsection{\label{subsec:renorm_algo}Renormalization algorithm}

With the concept of the constituting tiles, we sketch the outline of
the QCTMRG algorithm:

\begin{enumerate}
\item{\bf Construction of initial tiles}\\ 
In the special case of the Sirker-like decomposition \cite{Sir02},
the basic tile $W$ (figure~\ref{f:basic_tile}) 
as well as the column-to-column tile $P_V$
are built by addition of two mutually rotated transfer matrices $\tau$.
\begin{figure}[htbp]
\begin{center}
  \includegraphics[width=0.75\columnwidth]{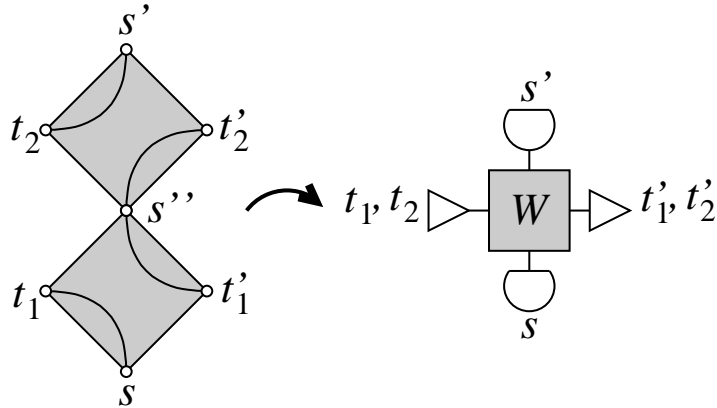}  
\end{center}
\caption{Graphical representation of the basic tile 
  $W$ in the Sirker-decomposition.  A composition of two transfer
  matrices can be depicted as a tile with four arms at the edges.  The
  shape of the arms symbolizes the symmetries of the tile and the
  basis of the underlying tensor operator.  }
\label{f:basic_tile}
\end{figure}
The left row-to-row tile $P_L$ and the left corner tile $C_L$ arise by
summing out the left free sites of $W$.  For the right row-to-row tile
$P_R$ and the right corner tile $C_R$ the right free sites of $W$ are
bent into the horizontal edges in order to aim a summation with the
neighboring site on the lower/upper tile when composing tiles.  In
figure\ \ref{f:initial_tiles}, the construction of all initial tiles
is depicted.  These initial tiles demand the construction of the
initial basis.  Starting from the one-site basis $\{\cats{\sigma}\}$,
the vertical basis of corner and column-to-column tile 
is\footnote{$\otimes$ denotes the usual tensor product.}
\mbox{$\basis{m_V} = \{\cats{\sigma_V}\} = \{\cats{\sigma} \otimes
  \cats{\sigma}\}$}, while the horizontal basis of left corner and
row-to-row tile is \mbox{$\basis{m_L} = \{\cats{\sigma_H}\} =
  \{\cats{\sigma}\}$}, and the horizontal basis of right corner and
row-to-row tile is \mbox{$\basis{m_R} = \{\cats{\sigma} \otimes
  \cats{\sigma}\}$} in the first iteration.

\begin{figure}[htbp]
  \begin{center}
    \includegraphics[width=0.95\columnwidth]{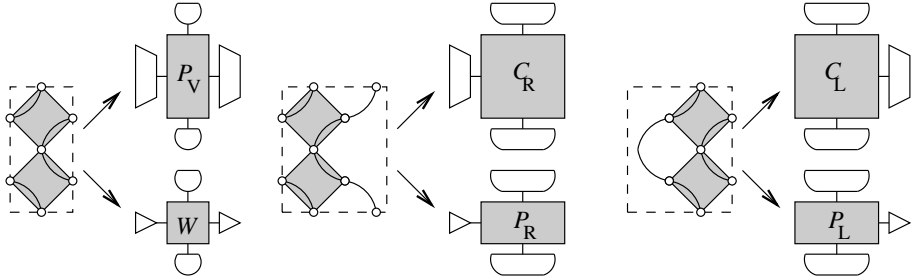}
  \end{center}
\caption{Graphical representation of the initial tiles.}
\label{f:initial_tiles}
\end{figure}

\item{\bf Calculation of expectation values}\label{it.calc}

  All tiles are combined to form a periodically closed two-dimensional
  lattice in order to determine the partition function of the system.
  Summing out the states on the inner edges, we obtain the desired
  partition sum $Z$ of the system.  The partition sum $Z(\Op)$ of the
  modified system with a certain operator $\Op$ situated in the middle
  of the lattice (see figure~\ref{f:qctmrg_measure}) can be realized
  similarly.  Thus, the thermodynamical expectation value
  $\big<\Op\big> = Z(\Op)/Z$ can be computed.
\begin{figure}[htbp]
\begin{center}
  \includegraphics[width=0.9\columnwidth]{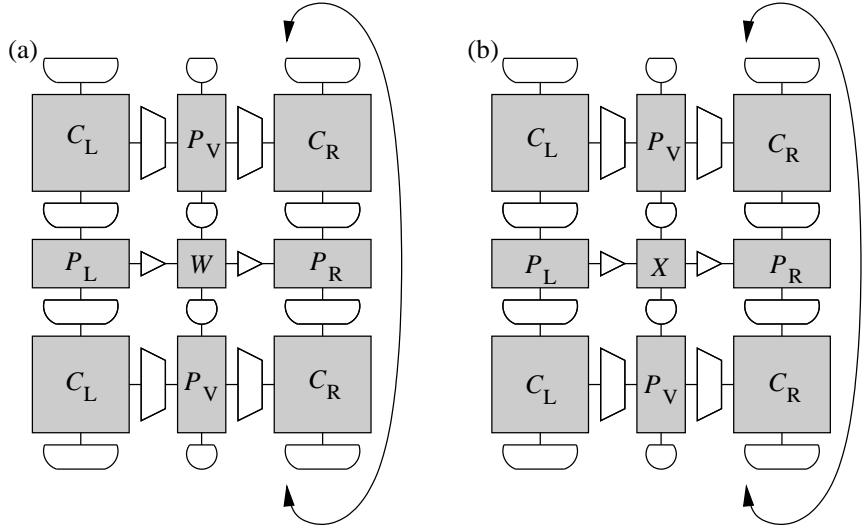}  
\end{center}
\caption{(a) Calculation of the partition function $Z$ and (b) of the 
  partition function of the system containing a certain operator
  $Z(\Op)$.}
\label{f:qctmrg_measure}
\end{figure}

\item{\bf Enlargement of system tiles}

  In the next step of the renormalization process, the system is
  expanded by enlargement of the corner tiles in applying a
  row-to-row, a column-to-column tile and the basic tile.
  Correspondingly, the row-to-row and column-to-column tiles have to
  be expanded by addition of the basic tile.  These enlargement steps,
  which technically correspond to matrix-matrix-multiplications, are
  illustrated in figure~\ref{f:enlarge_tiles}.  Enlargement of the
  tiles implies enlargement of the bases, which is done by simple
  tensor products $\basis{\tilde{m}_V} = \big\{\cats{m_V} \otimes
  \cats{\sigma_V}\big\}$, $\basis{\tilde{m}_L} = \big\{\cats{m_L}
  \otimes \cats{\sigma_H}\big\}$ and $\basis{\tilde{m}_R} =
  \big\{\cats{m_R} \otimes \cats{\sigma_H}\big\}$.  Note that the
  sizes of the bases grow by a factor of $4$ to $16$ depending on the
  underlying spin system, which correspondingly increases the size of
  the tiles.
\begin{figure}[htbp]
  \begin{center}
    \includegraphics[width=0.95\columnwidth]{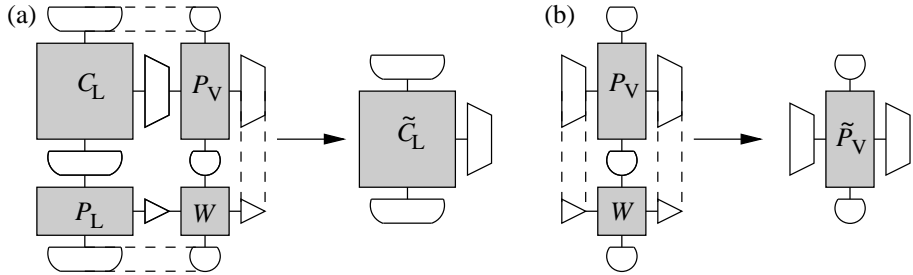}
  \end{center}
\caption{\label{f:enlarge_tiles} Enlargement of system tiles.
  (a) The enlarged corner tile $\tilde{C}_L$ is composed of the corner
  tile ${C}_L$, the column-to-column tile $P_V$, the row-to-row tile
  $P_L$ and the basic tile $W$.  (b) The enlarged the column-to-column
  tile $\tilde{P}_V$ is composed of the column-to-column tile $P_V$
  and the basic tile $W$.  }
\end{figure}

\item{\bf Construction of reduced density matrices}

  The crucial step in the DMRG-like renormalization procedure is the
  construction of the reduced density matrix.  In the Trotter
  decomposed lattice, three density matrices $\rho_V$, $\rho_L$ and
  $\rho_R$ are required in order to renormalize the three different
  types of bases appearing in the system (see
  figure~\ref{f:qctmrg_cuts}).  The concept of the reduced density
  matrix is to compose the corner matrices and to sum out all but one
  edge.  The picture of the reduced density matrices then corresponds
  to {\em cuts of the system} \cite{NO96,NO97}
\begin{figure}[htbp]
\begin{center}
  \includegraphics[width=0.95\columnwidth]{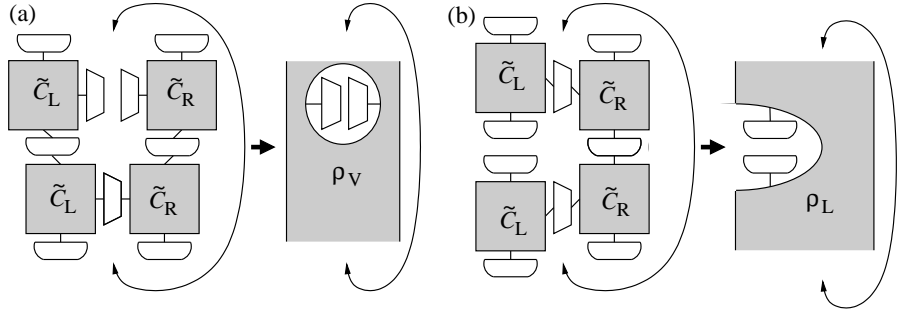}  
\end{center}
\caption{\label{f:qctmrg_cuts} Construction of the (a) vertical and (b) left
  reduced density matrix. Illustration as ``cuts of the system''.}
\end{figure}

\item{\bf Truncation of basis}

  The idea behind renormalization group procedures is to iteratively
  integrate out insignificant degrees of freedom.  In the context of
  DMRG-type algorithms, measuring the contribution of states for
  calculating the partition function is carried out by diagonalization
  of the reduced density matrix.  Those eigenstates with large
  eigenvalues will dominate because the partition function is nothing
  but the trace over the reduced density matrix.  Conservation of the
  bases' sizes demands truncation of the eigenstates with lowest
  weight. So, we establish a projection onto the $m$ states with
  largest eigenvalues as renormalization prescription.  Since a
  projection is obtained for each of the three bases, all edges of the
  system tiles are now reduced to a fixed size keeping only the most
  relevant states for calculation of the partition function.

\item{\bf Iteration}

Go to step (\ref{it.calc}) until desired system size is reached.

\end{enumerate}

\subsection{Normalization of growing tiles}

While our interest lies in the calculation of (local) expectation
values of a certain quantum mechanical system, we have to deal with
the partition functions of an iteratively increased classical system
in the QCTMRG-algorithm.  Thus, the partition function is a rapidly
growing entity leading to several huge matrix entries in the tiles'
numerical representation in each renormalization step.  So, the
program runs the risk of exceeding the numerical capacity of the
variables of the system.  To avoid this problem, a constant prefactor
is extracted in each renormalization step.

\subsection{Implementation}

Our implementation of the QCTMRG method follows the algorithm
presented in section \ref{subsec:renorm_algo}.  The ratio
$\epsilon=\beta/M$ is kept at a fixed value during the renormalization
procedure as in TMRG.  This provides, in contrast to TMRG, not only a
new (lower) temperature but also a different (larger) chain length
after each renormalization step.  Measurements of expectation values
can be made in the center of odd chains or at the two sites in the 
center of even chains.  Our implementation makes use of good quantum
numbers which drastically reduce computational effort.

\section{Results}\label{s:QCTMRG:Results}

As a first test, we compute the energy and the free energy of the
Ising model \cite{Isin25}.  The antiferromagnetic Ising chain $(J>0)$
with the classical Hamiltonian
\begin{equation}
  H = J \sum_{j} \sigma_{j} \sigma_{j+1}
\end{equation}
is exactly solvable by a transfer matrix method \cite{KrWa41}.  The
classical Ising spin $\sigma$ can take the values $+1,-1$.

\begin{figure}[tbp]
  \begin{center}
    \includegraphics[width=0.85\columnwidth]{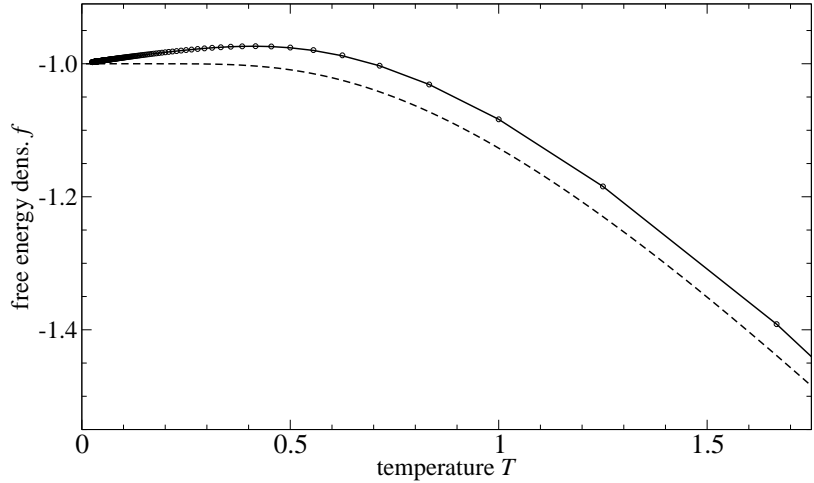}
  \end{center}
  \caption{\label{f.ising-f} Free energy per site vs.\ temperature 
    for the (classical) antiferromagnetic Ising chain of varying chain
    length.  The free energy (circles) from QCTMRG with $m=32$ and
    $\epsilon=0.05$ agrees to high precision to the exact value (full
    line).  Note that the chain length is $L=10/T$.  The free energy
    of the infinite chain is plotted (dashed line) for comparison.}
\end{figure}

\begin{figure}[tbp]
  \begin{center}
      \includegraphics[width=0.85\columnwidth]{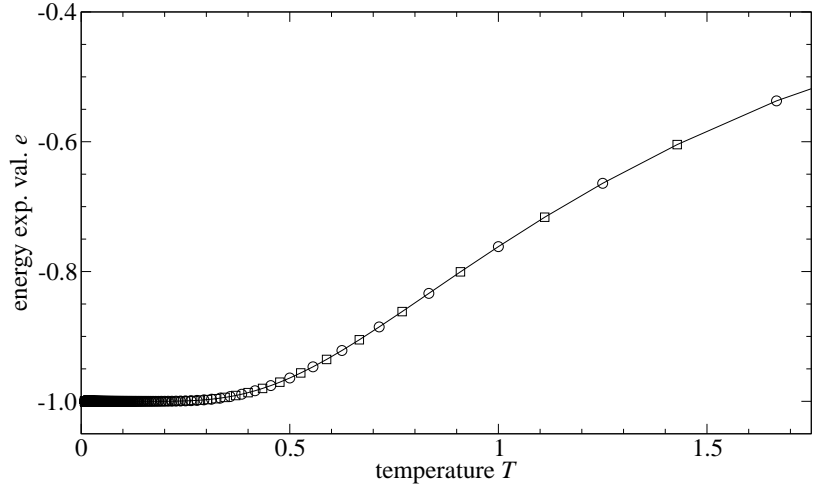}
  \end{center}
  \caption{\label{f.ising-e} Expected local energy  in the 
    center of the chain vs.\ temperature for the antiferromagnetic
    Ising chain of varying chain length.  The local energy expectation
    value in the center of the chain (circles and squares denote odd
    and even chain lengths) from QCTMRG with $m=32$ and
    $\epsilon=0.05$ agrees to high precision
    to the exact local energy (full line).
  }
\end{figure}

We have computed the thermodynamics of the Ising chain using the QCTMRG
algorithm, keeping $m=32$ states within the renormalization procedure.
The inverse factor of temperature $T$ and number of imaginary time 
steps $M$ has been chosen as $\epsilon=(TM)^{-1}=0.05$.  We expect
an excellent agreement with the analytical results because the Trotter
decomposition becomes exact for the classical Ising model.  In figures
\ref{f.ising-f} and \ref{f.ising-e} the numerical results are plotted
and, indeed, both data show the expected agreement.

We face a different situation in considering the antiferromagnetic
spin-$1/2$ Heisenberg chain
\begin{equation}
  H = J \sum_{j} \vec{S}_{j} \cdot \vec{S}_{j+1}.
\end{equation}
The model is exactly solvable \cite{B31} but quantum
fluctuations are no longer suppressed.  Thus, the Heisenberg chain can
serve as a non-trivial trial system for the QCTMRG algorithm.

\begin{figure}[htbp]
  \begin{center}
      \includegraphics[width=0.85\columnwidth]{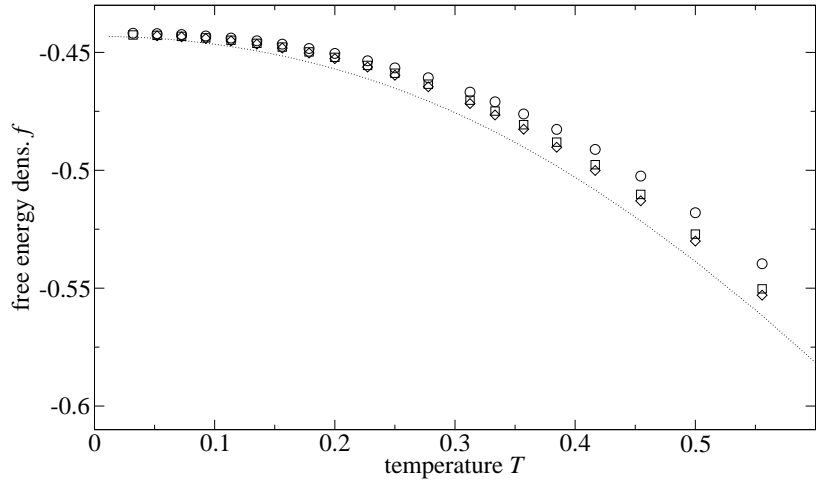}
  \end{center}
  \caption{\label{f.qctmrg-heisenberg-f} Free energy density 
    vs.\ temperature for the Heisenberg chain of varying chain length.
    The free energy density is plotted for different QCTMRG
    calculations (stolid lines) with a preserved number of states
    $m=50$ (circles), 150 (squares), 400 (diamonds).  Note that the
    chain length is $L=10/T$.  The free energy density (data from
    TMRG, $m=200$) of the infinite chain is plotted (dotted line) for
    comparison.}  
\end{figure}

\begin{figure}[htbp]
  \begin{center}
      \includegraphics[width=0.85\columnwidth]{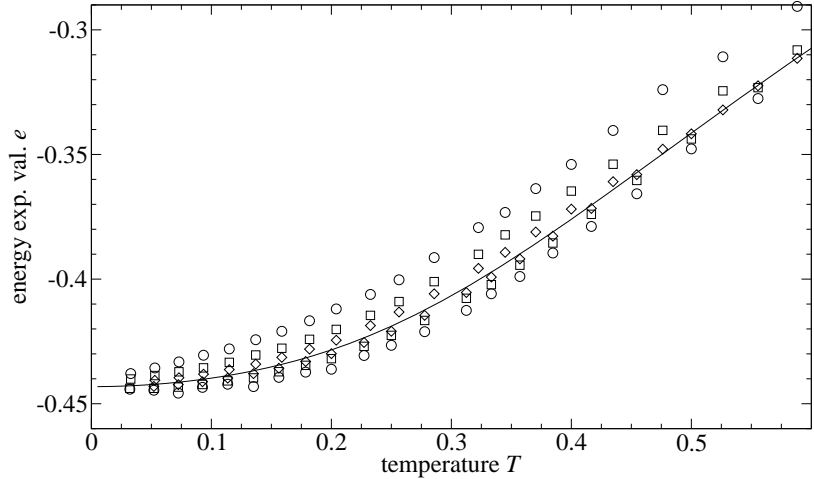}
  \end{center}
  \caption{\label{f.qctmrg-heisenberg-e} Energy per site
    vs.\ temperature for the Heisenberg chain of varying chain length.
    The energy expectation value at the chain center is plotted for
    different QCTMRG calculations with a preserved number of states
    $m=50$ (circles), 150 (squares), 400 (diamonds).  The upper points
    belong to systems with an odd number of sites, the lower points
    belong to systems with an even number of sites.  Note that the
    chain length is $L=10/T$.  The exact energy density of the
    infinite chain [22] is plotted as a full line for
    comparison.}  
\end{figure}

We calculated the free energy and the expectation value of the energy
operator in the middle of the chain, see
Figs.~\ref{f.qctmrg-heisenberg-f},~\ref{f.qctmrg-heisenberg-e}.
The calculations have been done with fixed ratio $\epsilon=0.05$ while
the preserved number of states $m$ from the renormalization was varied
from $m=50$ to $m=400$.

In figure~\ref{f.qctmrg-heisenberg-f}, the free energy density is
plotted against temperature.  Note that the chain length is related to
temperature by $L=10/T$.  With increasing $m$ the free energy tends to
converge.  Interestingly, the convergence is faster for lower values
of $T$ and correspondingly larger system sizes.  We account for that
point later in Section~\ref{subs:QCTMRG_challenges}.  For a
comparison, we added the well-converged data for an infinite chain
calculated by conventional TMRG.  We see the deviation of the QCTMRG
data from this curve more pronounced at higher temperatures which are
related to smaller system sizes.  Thus, we can interpret the deviation
as finite size effect.

\begin{figure}[htbp]
  \begin{center}
      \includegraphics[width=0.85\columnwidth]{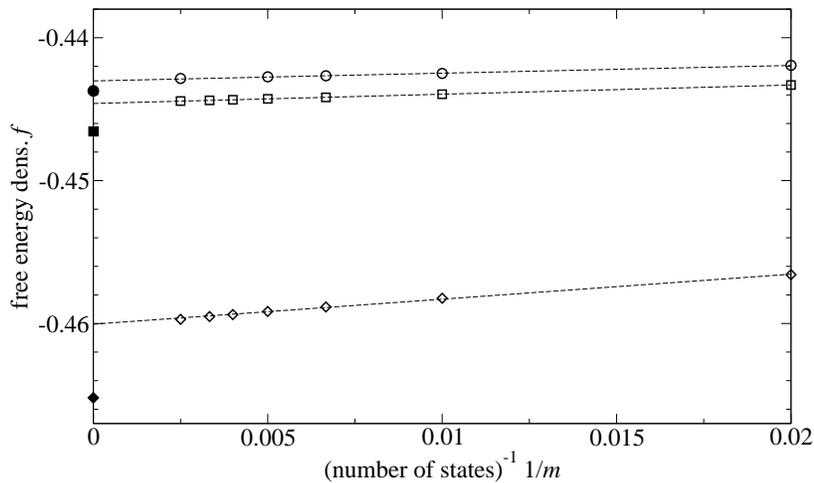}
  \end{center}
  \caption{\label{f.qctmrg-heisenberg-fvar} Free energy 
    density vs.\ inverse number of states for the Heisenberg chain of
    various lengths at various temperatures.  The free energy
    densities of chains with $T=0.25J$ and $L=40$ (diamonds), $T=0.1J$
    and $L=100$ (squares), and $T=0.04J$ and $L=250$ (circles) have
    been calculated by QCTMRG (open symbols) with a varying number of
    states kept during the renormalization.  The dashed lines are
    linear fits.  The filled symbols denote the free energy of an
    infinite system calculated by TMRG.
  }
\end{figure}

For three points of fixed temperature and chain length the free
energy density was extrapolated, see
figure~\ref{f.qctmrg-heisenberg-fvar}.  We find $f$ growing with $1/m$
for each point.  This results from the fact that the partition sum is
underestimated for smaller number $m$ of states kept.  This leads to a
monotonically increasing free energy for growing $1/m$.  As already
noticed earlier, we find better convergence for smaller temperature
and larger system sizes.

\begin{figure}[htbp]
  \begin{center}
    \includegraphics[width=0.95\columnwidth]{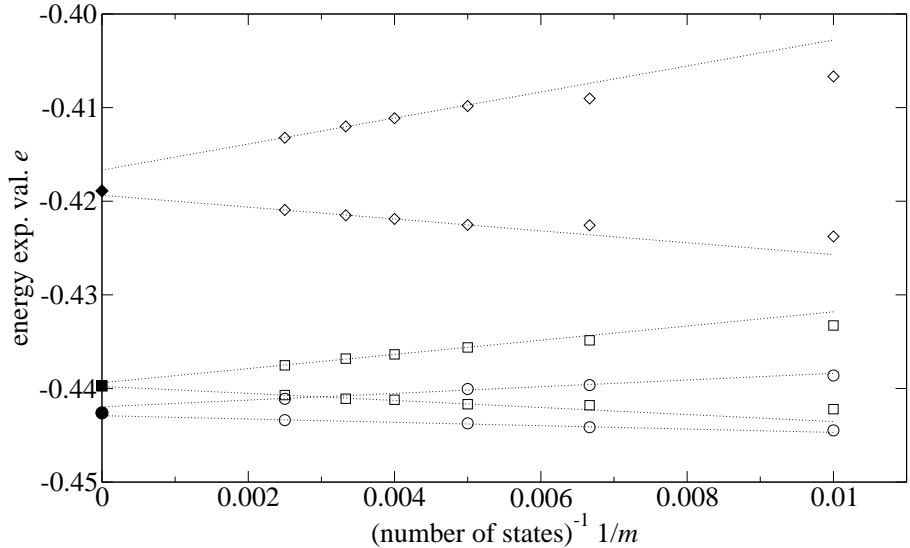}
  \end{center}
  \caption{Energy per site vs.\ inverse number of states for the 
    Heisenberg chain of various lengths at various temperatures.  The
    energy expectation values at the center of even chains (lower
    symbols) with $T=0.25J$ and $L=40$ (diamonds), $T=0.1J$ and
    $L=100$ (squares), and $T=0.04J$ and $L=250$ (circles) and of odd
    chains (upper symbols) with $T=0.256J$ and $L=39$ (diamonds),
    $T=0.101J$ and $L=99$ (squares), and $T=0.0402J$ and $L=249$
    (circles)  have been calculated.  We denote the exact energy
    density of the infinite chain by full symbols.  The dashed lines
    are linear fits.
  }
  \label{f.qctmrg-heisenberg-evar}
\end{figure}

The energy expectation value in the center of the chain has been
calculated for various system sizes and temperatures in
figure~\ref{f.qctmrg-heisenberg-e}.  Again we have $\epsilon=0.05$
which leads to a fixed relation $LT=10$ for the QCTMRG data.  Odd and
even chain lengths are included in this calculation, in contrast to
the free energy data which have only been given for even chains.  For
odd chains, the expectation value was taken from a plaquette at the
exact center of the decomposition.  For even chains, we considered one
of the two central plaquettes.  Even chain lengths appear as the
natural choice in the renormalization procedure.  Chains with an odd
number of sites have a well-defined central site which might give
better results.

We find the energy expectation value to lie for even systems below 
and for odd systems above the infinite-system value.  
The low-energy spectrum of the antiferromagnetic
Heisenberg chain involves spin-1/2 spinons which can be identified
with quantum domain walls.  The ground state is a total spin singlet
on chains with an even number of spins.  In odd chains there is no
spin singlet ground state and so always a spinon ``excitation'' is
present.  In the domain wall picture, there is always a kink present.
For this reason, we expect the chains with an even number of sites to
possess a lower local energy than chains with an odd number of sites.
However, we expect that the expectation values for even and odd chains
converge to the same limit in an infinite chain.  This description
agrees well with our observed behaviour.

In either case, we have a strong dependence of the expectation value
on the number of preserved states $m$ in the renormalization group.
For both even and odd number of sites, we can still distinguish the
data curves up to low temperatures even for high values of $m$.  Yet,
the convergence for even system sizes is faster than for odd sizes.

To get a more quantitative picture we plotted the convergence of
several points of fixed temperature and system size, see
figure~\ref{f.qctmrg-heisenberg-evar}.  Like in the free energy data,
the high-$T$ and larger sized systems show a better convergence.  We
included linear fits for a better understanding.  The expectation
values for even/odd sized system at finite temperatures seem to differ
even in an $1/m$ extrapolation. The infinite-chain expectation value
lies well-between those boundaries.  For larger system sizes and lower
temperatures we obtain a better convergence in $1/m$.

The results from this section shall serve as an illustrating
background for a discussion of the scope of application and the limits
of the QCTMRG technique.

\section{Discussion of the QCTMRG method}\label{s:QCTMRG:Discussion}

A crucial attribute for the QCTMRG algorithm is the fixed relation
between system size and temperature during the renormalization loops.
This condition sets the focus of the method to finite size systems, in
contrast to the TMRG.  Yet, a significant advantage in computational
resources would make the QCTMRG an interesting choice for
low-temperature studies of very large systems where finite-size
effects play a marginal role.

\subsection{Running time and storage use}

For a view on the algorithm's running time and storage use, we studied
the scaling behaviour with the number of states kept $m$ of the
different steps in the algorithm.  We find the algorithm to take
asymptotically $\mathcal{O}(m^4)$ elementary floating point operations
(FLOP) and a storage use of order $\mathcal{O}(m^3)$ floating point
numbers for large $m$.  However, most of the time is spent in
matrix-matrix-multiplications operations where a naive approach would
give a running time of $\mathcal{O}(N^3)$ FLOP for multiplication of
two $N\times N$ matrices.  In contrast to this, the fastest algorithm
currently known has an asymptotic running-time of
$\mathcal{O}(N^{2.376})$ FLOP \cite{CoWi90}.  So we expect the
running-time to scale asymptotically with well-below fourth order in
$m$ in a clever implementation.

In case of the TMRG algorithm, the by-far most time consuming part is
finding the largest eigenvalues of the transfer-matrix.  We, thus,
expect the asymptotic scaling behaviour to be dominated by the
employed Arnoldi method which performs with between $\mathcal{O}(1)$
and $\mathcal{O}(N)$ $\times$ the cost of a certain matrix-vector
product when $N\times N$ is the size of the matrix.  In our case, the
necessary matrix-vector product has a cost of $\mathcal{O}(m^3)$ FLOP
and the total matrix has a basis of size $N \propto m^2$.  We thus
expect the total running-time to vary between $\mathcal{O}(m^5)$ worst
case and $\mathcal{O}(m^3)$ best case.  Further reduction might be
achieved when clever algorithms for matrix multiplications are
implemented.  The storage scales with the size of the
system-block-transfer-matrix ($\mathcal{O}(m^2)$ numbers).

From the theoretical point of view, the QCTMRG and the TMRG algorithms
have similar running-times in the large-$m$ limit.  The TMRG algorithm
will be favorable in those cases where the Arnoldi method reaches a
fast convergence.  For some ill-posed problems, however, the QCTMRG
algorithm might have an advantage.  The TMRG algorithm is the clear
winner when storage usage is a sensible quantity.

In our implementation of both methods, we observed a roughly similar
running time of both algorithms for different number of preserved
states $m$.  The storage use was, indeed, higher in the QCTMRG
algorithm.

\subsection{Fundamental challenges}\label{subs:QCTMRG_challenges}

Yet, we still face some fundamental challenges of the QCTMRG algorithm
in addition to the advantages of TMRG concerning the use of system
resources.  As already mentioned above, the quantum character of the
system enforces periodical boundary conditions in Trotter direction
which makes the two-dimensional corner transfer-matrix from Nishino's
CTMRG a three-dimensional tensor in the case of QCTMRG.  This is the
reason behind an increase in running-time and storage which is crucial
and partly destroys the advantages from the development of CTMRG over
TMRG.

Another more subtle point involves the spectra of the reduced density
matrices from the renormalization procedure.  Consider the limit $T\to
\infty$ which is close to the starting point of TMRG and QCTMRG where
$\beta=1/T=\epsilon M$ is small.  Here, the local transfer-matrix of
the Trotter decomposition reduces to
\begin{eqnarray}
\bras{{s'\!\!}_{i} {s'\!\!}_{i+1}} \tau \cats{s_{i} s_{i+1}} 
:= \bras{{s'\!\!}_{i} {s'\!\!}_{i+1}}
e^{-\epsilon h_{i,i+1}} \cats{s_{i} s_{i+1}}
= \delta_{{s'\!\!}_{i},s_{i}} \delta_{{s'\!\!}_{i+1},s_{i+1}}
\end{eqnarray}
which means that the initial spin configuration will not be changed by
the transfer-matrix.

Now, we consider a Trotter decomposition for infinite temperature or
vanishing $\epsilon$.  The chosen graphical representation depicts the
Kronecker symbols as lines passing through the transfer-matrix
plaquettes.  If the system is periodically closed in Trotter direction
and has open boundaries in space direction, we can imagine the paths
as non-interacting lines around a cylinder.  This depicts the trace
from the partition sum.

Building a reduced density matrix introduces a vertical or horizontal
cut into the system.  Consider the case of a cut in Trotter direction
which corresponds to the cut in TMRG.  One ``spin path'' has been cut
through while all the others stay intact, {\it i.e.} they still are
summed out in the reduced density matrix.  Only the intersected thread
determines the eigenspectrum of the reduced density matrix.  With a
straightforward calculation we find one or two degenerate eigenvalues
depending on whether we intersect an odd or even number of sites.  All
other eigenvalues remain zero in this case.  This turns the
renormalization step, which is a truncation of basis states, to be
highly effective.  In a system with small $\epsilon$ we still find a
small number of dominating eigenvalues which lead a DMRG
renormalization to success.

The opposite situation is faced when the cut is made in space
direction.  If the underlying spin chain has $2L$ spins,
the density matrix will ``cut'' as much as $L$ paths.  All $L$ spins
act separately and, thus, all spin configurations have the same
contribution to the partition function.  Merely the remaining degrees
of freedom will be summed out.  Consequently, we are left with a
reduced density matrix which has $S^L$ degenerate eigenstates when $S$
is the spin size.  No effective truncation can be found.  The
situation in cases with high, but finite temperature is certainly less
ill-posed.  Though, still we expect a slow decay of the spectrum of
the reduced transfer matrix.

This scenario explains why the calculated QCTMRG data shows a most
significant deviation from the expected values at higher temperatures.
Starting with a small system size the algorithm can still handle the
system with high precision.  For larger system sizes, a truncation has
to be done during the renormalization step which cannot be optimal
since the reduced density matrix will still have a flat spectrum.  At
even larger system sizes and lower temperatures the spectrum of
horizontal density matrix will become acceptably well-behaved again.

\section{Conclusions and Outlook}

We developed a new method for finite-temperature studies of
one-dimensional quantum systems based on the CTMRG of Nishino.  The
free energy densities and thermal energy expectation values at the
chain centers have been successfully calculated for the classical
Ising chain and the antiferromagnetic spin-$1/2$ Heisenberg chain by
the Quantum Corner-Transfer Matrix DMRG.  Reliable results were given
for finite temperatures and system sizes.  Yet, the algorithm faces
two difficulties:
\begin{itemize}
\item Periodic boundary conditions reduce the efficiency.
\item The reduced density matrix in space direction has a slowly
  decaying eigenspectrum.
\end{itemize}
If finite-temperature data in the thermodynamic limit are aimed at, the
quantum TMRG method is certainly still the method of choice.  At least
one of the mentioned problems should be solved to make the QCTMRG
technique an attractive option.


\paragraph{Acknowledgments}
We dedicate this paper to Dietrich Stauffer on the occasion of his
retirement.
This work has been performed within the research program SFB
608 of the \emph{Deutsche Forschungsgemeinschaft}.


\end{document}